\documentclass[aps,pre,twocolumn,superscriptaddress,showpacs]{revtex4}

\usepackage{amsmath}  \usepackage{amsfonts} 
\usepackage{graphicx} \usepackage{color} \usepackage{mathrsfs} 
\usepackage{subfigure} \usepackage{color} \usepackage{ulem} \usepackage{siunitx} \usepackage{relsize}
\usepackage{multirow} \usepackage{bigints}     
\DeclareMathOperator{\sinc}{sinc}

\begin{document}

\title{Electromagnetic scattering by a partially graphene-coated dielectric cylinder : efficient computation and multiple plasmonic resonances}

\author{Youssef Jeyar}
\affiliation{ Laboratoire Charles Coulomb (L2C), UMR 5221 CNRS-Universit\'{e} de Montpellier, F-34095 Montpellier, France}

\author{Mauro Antezza}
\affiliation{ Laboratoire Charles Coulomb (L2C), UMR 5221 CNRS-Universit\'{e} de Montpellier, F-34095 Montpellier, France}
\affiliation{Institut Universitaire de France, 1 rue Descartes, Paris Cedex 05 F-75231, France}

\author{Brahim Guizal}
\affiliation{ Laboratoire Charles Coulomb (L2C), UMR 5221 CNRS-Universit\'{e} de Montpellier, F-34095 Montpellier, France}
\date{\today}
\begin{abstract}
We present a numerical approach for the solution of electromagnetic scattering from a dielectric cylinder partially covered with graphene. It is based on a classical Fourier-Bessel expansion of the fields inside and outside the cylinder to which we apply ad-hoc boundary conditions in presence of graphene. Due to the singular nature of the electric field at the edges of the graphene sheet, we introduce auxiliary boundary conditions. The result is particularly simple and very efficient method allowing the study of diffraction from such structures. We also highlight the presence of multiple plasmonic resonances that we ascribe to the surface modes of the coated cylinder. 
\end{abstract}

\maketitle
\section{Introduction} \label{I}
Scattering of electromagnetic waves from a dielectric or a metallic circular cylinder is rather a simple and classical problem \cite{book1}. In this situation, the incoming and outgoing fields can be represented in terms of cylindrical waves expressed through Fourier-Bessel expansions and all the channels are independent because of the circular symmetry and homogeneity of the cylinder. Usually, when the wavelength of the incoming wave is much larger than the diameter of the cylinder (this is the {\it subwavelength regime}), the wave is barely scattered. Interestingly, it has been shown that when cylindrical structures involve plasmonic materials (noble metals or graphene) they are able to exhibit quite unusual phenomena such as superscattering \cite{Shanhui, CompactSS} ({\it i.e.} they scatter light as if they were much larger than their actual size) and/or invisibility  ({\it i.e.}  they scatter light as if they were much smaller than their actual size). Furthermore when the cylinder is partially covered with a PEC (Perfect Electric Conductor) circular strip \cite{GuizalPRE} or a circular strip of graphene \cite{Integral1, Integral2}, this breaks the symmetry and homogeneity at the level of the surface and then all the channels can be mixed up leading to much richer physical behaviour.  In \cite{Integral1} an integral equation approach has been used to explore the interplay between plasmonic-resonances and photonic-jet effects in THz wave scattering by a graphene-covered dielectric cylinder.  The same approach has been used in \cite{Integral2} to study the performances of THz antenna made of a circular dielectric rod with conformal strip of graphene. 

On the other hand, an interesting and particularly simple approach to study scattering from partially covered circular cylinders is to use the periodicity of the conductivity function and introduce its Fourier expansion directly in the boundary conditions. After projection on the Fourier basis, this leads to an algebraic system linking the outgoing amplitudes of the fields to the incoming ones. This approach will be called the FMM (Fourier Modal Method) since it is the counterpart (or an extension) of the well known namesake method introduced for planar strips gratings \cite{GUIZAL1999, Khavasi}. In this very context (that of strip gratings) this method works extremely well in the case of transverse electric polarization (TE : the electric field is parallel to the direction of invariance of the grating), but, unfortunately, may face serious convergence problems in the case of transverse magnetic polarization (TM : the magnetic field is parallel to the direction of invariance of the grating). This is due to the fact that the tangential component of the electric field at the edges of the graphene sheet is null, hence preventing from using the correct Fourier Factorization rules \cite{Li:FFF,Khavasi} (cf. subsection II-B ). In \cite{Khavasi} A. Khavasi proposed a strategy to improve the situation through the introduction of Approximate Boundary Conditions (ABC) allowing the use of the correct Fourier Factorization rules. This approach (that we will call FMM-ABC) shows a certain efficiency but do not completely fix the problem of convergence, especially for structures involving sharp resonances. Moreover, the technique introduces a new free parameter whose tweaking is very delicate \cite{Taiwan_LBF}. It is also important to emphasize that the tangential electric field is singular in the vicinity of the graphene strips edges and is at the origin of the slow convergence for both the classical FMM and the FMM-ABC.   
Very recently, an alternative approach has been put forward by R. B. Hwang \cite{Taiwan_LBF} in order to solve this issue.  It is based on the use of a supplementary expression of the tangential electric field, right at the level of the interface, under the form of a special expansion in terms of Local Basis Functions (LBF) able to reproduce the aforementioned singularities (hereafter, this method will be called FMM-LBF). This proved to be extremely efficient, not only from the standpoint of convergence of the efficiencies but also in the representation of the field around the graphene strips.   

In this work, we introduce (in section II) the FMM in the context of scattering by a dielectric cylinder partially covered with graphene as well as its FMM-ABC and FMM-LBF extensions. In section III, we examine the convergence and stability of these three approaches and show the superiority of the FMM-LBF. Finally, we exploit the latter to highlight the presence of multiple resonances in the scattering efficiency spectrum for partially graphene-covered dielectric cylinders (on the contrary of what is observed for fully and homogeneously graphene-covered cylinders) and link them to the plasmonic surface modes over the structure.

\section{Theoretical framework}\label{II}
\begin{figure}[ht!]
\centering
{\includegraphics[width=8cm]{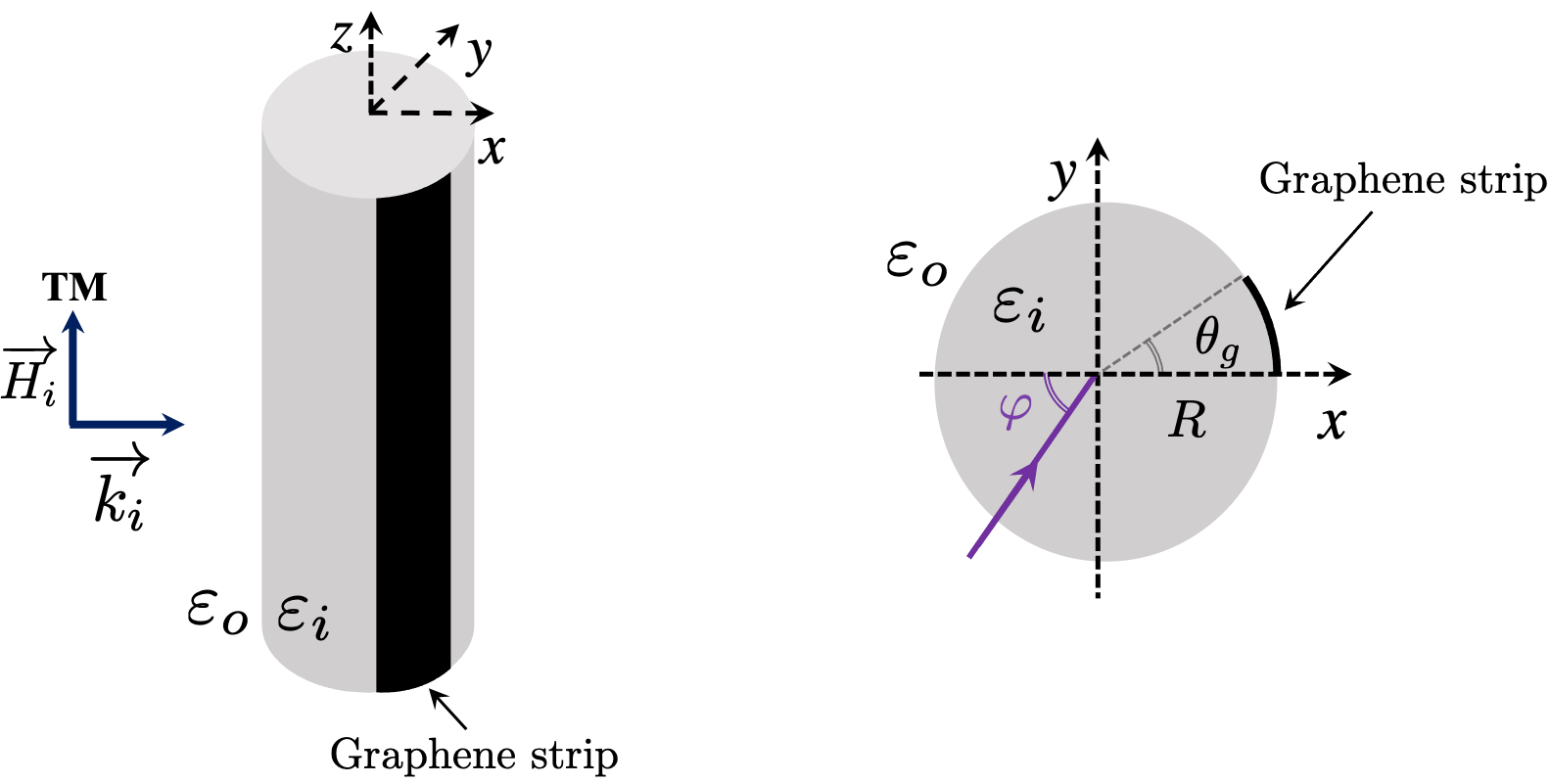}}
\caption{Sketch of the diffraction problem under consideration: an electromagnetic plane wave hits the covered dielectric cylinder under classical diffraction {\it i.e.} the incident wave vector is perpendicular to the direction of invariance $oz$.}
\label{Figure1}
\end{figure}

The physical problem under study is depicted in figure \ref{Figure1} where a Transverse Magnetic (TM : the magnetic field is parallel to the direction of invariance $oz$) linearly polarized electromagnetic plane wave, with vacuum wavelength $\lambda$,  illuminates a dielectric cylinder (radius $R$ and relative dielectric permittivity $\varepsilon_i$) under classical incidence (the incident wave vector is perpendicular to $oz$) with an angle $\varphi$. The cylinder lies in a host medium (relative dielectric permittivity $\varepsilon_o$) and may be covered with graphene strips whose electromagnetic behaviour is captured through their surface optical conductivity $\sigma(\omega)$. Both media are supposed non magnetic (relative magnetic permeabilities $\mu_{i/o} = 1$). In the following we will use harmonic Maxwell's equations with the time convention $e^{-i\omega t}$.  

Following reference \cite{Graphene1}, the graphene conductivity $\sigma(\omega)$ can be written as a sum of an interband and an intraband contributions, respectively given by :
 \begin{equation}  \label{eq1}
\begin{array}{ll}
 \sigma_R (\omega) = \dfrac{i}{\omega + i\Gamma} \dfrac{2e^2 k_B T}{\pi \hbar^2 } \ln \left(2 \cosh \dfrac{\mu}{2 k_B T} \right)\\ \\
  \sigma_I (\omega)= \dfrac{e^2}{4\hbar}\left[G \left( \dfrac{\hbar \omega}{2}\right) + i \dfrac{4\hbar \omega}{\pi}\mathlarger{\int}_0^{+\infty} \dfrac{G(\xi)-G(\frac{\hbar \omega}{2})}{(\hbar \omega)^2 - 4\xi^2} d\xi \right]
   \end{array} 
 \end{equation} 
$G(x) = \sinh (x/k_B T) / [\cosh (\mu/k_B T) + \cosh(x/k_B T)]$, T is the graphene strip temperature, $\Gamma = 2\pi/\tau$, $\tau$ being the relaxation time and $\mu$ the chemical potential.

In cylindrical coordinates, the $z$ components of magnetic fields in the inner and outer media can be written in terms of the elementary Fourier-Bessel solutions \cite{book1} : 
\begin{equation}  \label{eq2}
\begin{cases}
H_{zo} (r,\theta) = \sum\limits_{n} \left\{ a_n J_n (k_o r ) + b_n H_n ^+(k_o r)\right\}e^{in\theta}\\
H_{zi} (r,\theta) = \sum\limits_{n} c_n J_n (k_i r)e^{in\theta}
 \end{cases} 
 \end{equation}
where $k_{i/o}=k_0n_{i/o}$  with $k_0=2\pi/\lambda$ and $n_{i/o}=\sqrt{\varepsilon_{i/o}}$,  $J_n$ and $H_n^+$ being the $n$-th order Bessel and Hankel functions of the first kind, respectively ($n\in \mathbb{Z}$). Here $a_n = (i e^{-i\varphi})^n$ are the coefficients of the incident wave, while $b_n$ and $c_n$ are the scattering coefficients to be determined and from which all the physical quantities of interest can be readily computed. In the following, we will be mostly interested by the scattering efficiency,  the expression of which is given by: 
 \begin{equation}  \label{eq3}
 Q_{s} = \dfrac{2}{k_o R} \sum_n  |b_n|^2
 \end{equation}
The determination of $b_n$ and $c_n$ is accomplished via the boundary conditions at $r=R$ : $H_{zo} (R,\theta)-H_{zi} (R,\theta) =\sigma E_{\theta i}(R,\theta)$ and $E_{\theta i}(R,\theta)=E_{\theta o}(R,\theta)$, for all $\theta \in [0,2\pi]$, where $E_{\theta i/o} = (-i Z_0/k_0  \varepsilon_{i/o}) \partial_{r}H_{z i/o}$ ($Z_0$ being the impedance of vacuum). Once projected on the Fourier basis $(e^{in\theta})_{n\in \mathbb{Z}}$ \footnote{Throughout the paper, we use the classical inner product : $\langle f,g \rangle = \frac{1}{2\pi} \int_0^{2\pi} f(\theta)g^*(\theta)d\theta$, the star indicating complex conjugation.}, these equations will furnish a set of algebraic equations linking the unknown coefficients to those representing the incident field.      

When the cylinder is fully and homogeneously covered with graphene, applying the aforementioned procedure leads to the following simple formulas of the scattering coefficients : 
\begin{equation}\label{eq4}
\begin{cases}
b_n= \dfrac{n_i J_{ni} J'_{no}-n_oJ_{no}J_{ni}'+i \eta J'_{ni}J'_{no}}{n_o H_n^+J'_{ni}-n_i J_{ni}H_n^{+'}-i\eta J'_{ni} H_n^{+'} }a_n \\
c_n = \dfrac{n_i}{n_o}\dfrac{1}{J'_{ni}} \left\{ J'_{no} a_n + H^{+'}_n b_n \right\}
\end{cases}   
\end{equation}
Where $J_{n(i/o)}=J_{n}(k_{i/o}R)$, $J_{n(i/o)}'=J_{n}'(k_{i/o}R)$, $H^+_{n} = H^+_{n}(k_{o}R)$,  $H^{+'}_{n} = H^{+'}_{n}(k_{o}R)$, $\eta=Z_0 \sigma(\omega)$ and where the primes denote the derivation with respect to variable $r$.  

\subsection{The classical FMM}\label{II-1}
If the cylinder is partially covered with graphene, one can use the classical FMM \cite{GUIZAL1999, GuizalPRE} where the conductivity function is expanded into Fourier series with respect to variable $\theta$ $\left(\sigma(\theta)= \sum_p \sigma_p e^{ip\theta}\right)$, and inserted into the boundary conditions, which leads to the following expressions for the scattering coefficients:
 \begin{equation}
 \left\{
\begin{array}{ll}\label{eq5} 
b = B^{-1} \hspace{0.01cm}  A \hspace{0.05cm} a \\
 c = \dfrac{n_i}{n_o} J_{i} ^{'-1}\left\{ J'_{o} \hspace{0.05cm} a +  H^{+'}\hspace{0.05cm} b   \right\}
   \end{array}\right.  
\end{equation}
where matrices $A$ and $B$ are given by : 
\begin{equation}
 \left\{
\begin{array}{ll}\label{eq6} 
   A =  n_i  J_{i}J'_{o}- n_oJ_{o} J'_{i}  +i Z_0 J'_{i}\left\|\sigma\right\|J'_{o}\\
   B = n_o H^{+} J'_{i}- n_i J_{i}H^{+'} - i Z_0 J'_{i} \left\|\sigma\right\| H^{+'}  
   \end{array}\right.  
\end{equation}
Where we introduced vectors $a, b, c=(a_n, b_n, c_n)_{n \in \mathbb{Z}}$ and diagonal matrices $J_{i/o} = diag(J_{n(i/o)})$, $H^{+}=diag(H_n^{+})$, $J'_{i/o} = diag(J'_{n(i/o)})$, $H{'^+}=diag(H_n{'^+})$. $\left\|\sigma\right\|$ is the Toeplitz matrix built from the Fourier components of the conductivity such that $\left\|\sigma\right\|_{mn} = \sigma_{m-n}$. 
A rapid comparison between expressions in equations (\ref{eq4}) and (\ref{eq5}-\ref{eq6}) shows that they have the same structure where the role of $\eta$ in the former equations is played by $Z_0\left\|\sigma\right\|$ in the second set of equations. In the former case, the scattering channels are independent of each other whereas in the second case they are coupled through the matrix $\left\|\sigma\right\|$ (or the graphene strip, actually).   
\subsection{The FMM with ABCs} \label{II-2}
\begin{figure}[ht!]
\centering
{\includegraphics[width=8cm]{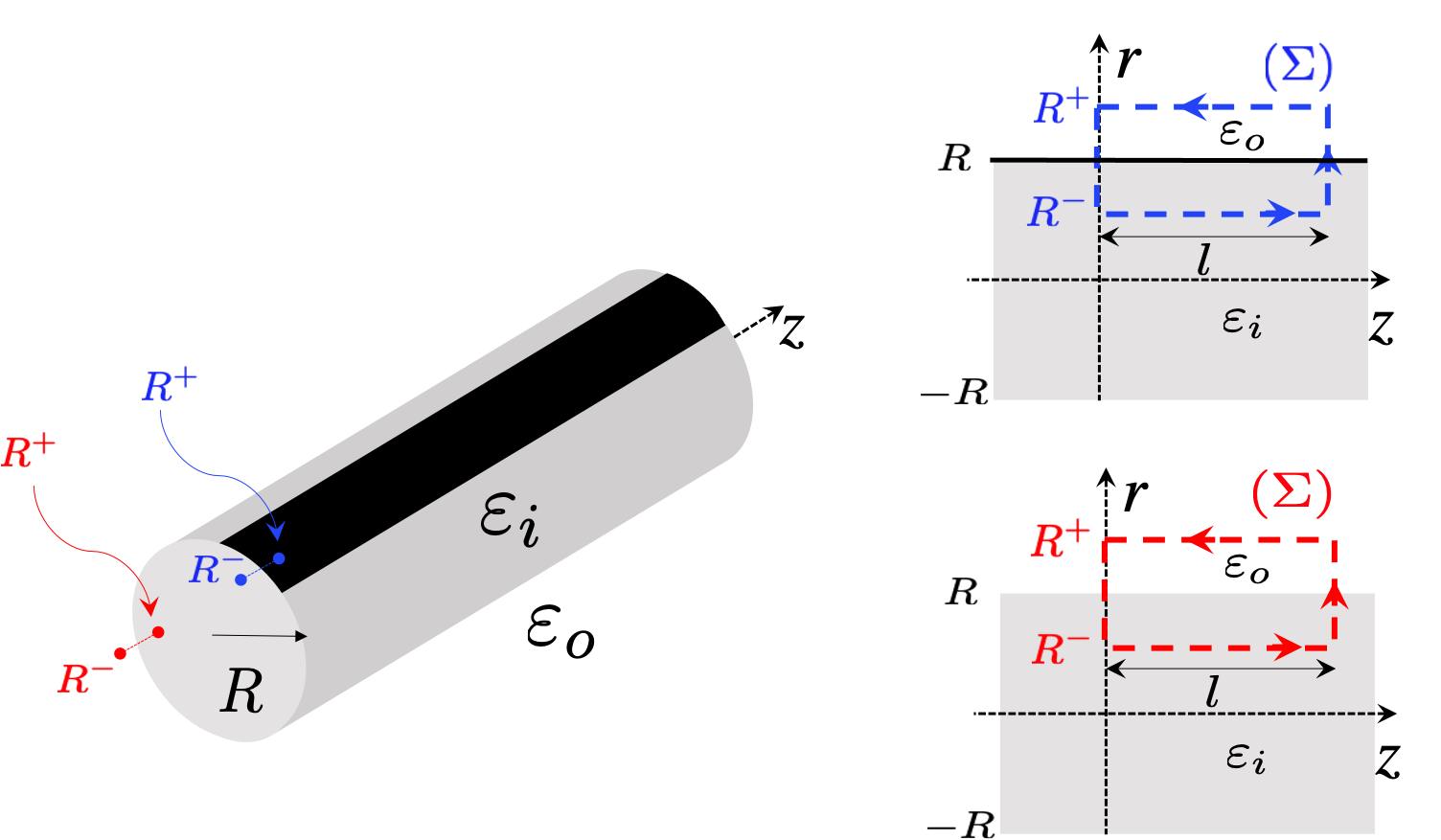}}
\caption{Sketch of the contours used to derive the Approximate Boundary Condition (ABC). Top-Right :  a contour across graphene and Bottom-Right : a contour on a bare part of the cylinder.}
\label{Figure2}
\end{figure}

In the classical FMM, Laurent's direct rule has been used to Fourier factorize the product $\sigma(\omega, \theta) E_{zi}(R,\theta)$ whereas it is the inverse rule \cite{GranetGuizal, Li:FFF} that should be used. Unfortunately, the inverse rule shall make use of the reciprocal of the conductivity function, $1/\sigma(\theta)$, whereas the latter takes infinite values on the part of the circumference without graphene. This clearly forbids the use of such a rule. To circumvent this problem, A. Khavasi, in the case of planar graphene strip gratings, introduced approximate boundary conditions \cite{Khavasi} leading to an effective conductivity whose reciprocal is never infinite, thus bringing back the possibility to use the inverse rule. Here we are facing the same situation and hence can transpose the work done in \cite{Khavasi} in cylindrical geometry. To be more specific, let us start from the Ampere's law and apply it to the closed rectangular loops $(\Sigma)$ shown in figure \ref{Figure2}. Then for a fixed $\theta$ :           
\begin{multline} \label{eq7} 
   l \left\{ H_{zi}(R^-,\theta)- H_{zo}(R^+,\theta) \right\}  \\  
                            = \int_{R^-}^{R^+} \int_0^l \left\{  \sigma(\theta) \delta(r-R) -i\omega \varepsilon\right\} E_{\theta}(r,\theta)dzdr 
\end{multline}

with $R^\pm = R \pm \Delta/2$, $\Delta$ and $l$ being the lengths associated to the $(\Sigma)$ contours. For sufficiently small $\Delta$, $E_{\theta}(r,\theta)$ can be approximated by $E_{\theta i}(R,\theta)$ (or $E_{\theta o}(R,\theta)$) and taken out of the integral, which leads finally to the ABC :  
\begin{equation} \label{eq8} 
   H_{zi}(R^-,\theta)- H_{zo}(R^+,\theta) = \widetilde{\sigma}(\theta)E_{\theta i/o}(R,\theta) 
\end{equation}

where $\widetilde{\sigma}(\theta) = \sigma(\theta) - i\omega \varepsilon_0 \Delta (\varepsilon_i+\varepsilon_o)/2$ is the effective surface conductivity (depending on the new parameter $\Delta$) that, now, never goes to zero.

It is now possible to use the inverse rule for this ABC which combined with the continuity of the electric field leads to the new expressions of the scattering coefficients that are exactly those of equation \ref{eq5} where the matrices $A$ and $B$ are replaced by :
\begin{equation}\label{eq9} 
 \left\{
\begin{array}{ll}
   A =  n_i  \widetilde{J}_{i}J'_{o}- n_o\widetilde{J}_{o} J'_{i}  +i Z_0 J'_{i}\left\|\widetilde{\sigma}^{-1}\right\|^{-1}|J'_{o}\\
   B = n_o \widetilde{H}^{+} J'_{i}- n_i \widetilde{J}_{i}H^{+'} - i Z_0 J'_{i} \left\|\widetilde{\sigma}^{-1}\right\|^{-1} H^{+'}  
   \end{array}\right.  
\end{equation}
where $ \widetilde{J}_{i} = diag(J_{ni}(k_iR^-))$, $ \widetilde{J}_{o} = diag(J_{no}(k_oR^+))$ and $ \widetilde{H}^{+} = diag(H_{n}^+(k_oR^+))$. \\
Although this approach has been shown to be effective (at least for the far field calculations) for planar diffraction gratings, it appears to experience some problems (as will be shown below) in the case of circular strips, especially near sharp resonances that are generally the most interesting spectral zones. 

\subsection{The FMM with LBFs}\label{II-3}
As said above, the FMM equipped with ABC can ensure convergence, but for that, a suitable choice of $\Delta$ is necessary which requires ancillary calculations. In addition, $\Delta$ depends on the wavelength. It is for this reason that a new and very efficient method has been proposed \cite{Taiwan_LBF} to model diffraction from planar strips and that we adapt here for cylindrical strips.\\
We start from the expressions of the fields in the inner and outer media (equations (\ref{eq2})) and add an expression of the electric field valid on the interface $r=R$, given in terms of the Local Basis Functions (LBFs), $g_m(\theta)$ and $s_m(\theta)$, that reproduce its singularities at $\theta=0$ and $\theta=\theta_g$:
\begin{equation}\label{eq10} 
 E_{\theta} (\theta)= \begin{cases}
\sum\limits_{m=1}^{N_g} p_mg_m(\theta) \quad,  0 \le \theta \le \theta_g \\ \\
\sum\limits_{m=0}^{N_s-1} q_ms_m(\theta) \quad,  \theta_g < \theta < 2\pi\\ 
 \end{cases}
 \end{equation}
where { \begin{center} $ \begin{cases}  
g_m(\theta)=\sin{(m\pi \theta/\theta_g)}\\ 
s_m(\theta)=\cos{(m\pi(\theta-\theta_g)/\bar{\theta}_g)/\sqrt{(\bar{\theta}_g/2)^2-(\theta-\theta_c)^2}}\\ 
\end{cases}$ \end{center} } 
$p_m$ and $q_m$ are the associated expansion coefficients, $\bar{\theta}_g=2\pi-\theta_g$ and $\theta_c=(\theta_g+2\pi)/2$. Parameters $N_g$ and $N_s$ represent the number of basis functions in the graphene strip and slit regions, respectively.\\

Now, for the boundary conditions, we must have : 
\begin{equation}\label{eq11} 
 \begin{cases}
E_{\theta i}(R,\theta)=E_{\theta o}(R,\theta) \\ 
H_{zo} (R,\theta)-H_{zi} (R,\theta) =\sigma(\theta) E_{\theta i}(R,\theta)\\ 
E_{\theta i}(R,\theta)=E_{\theta}(\theta) \\
\end{cases} 
\end{equation}
The first two equations correspond to the classical boundary conditions used before. As for the third one, it enforces the electric field, at $r=R$, to match the one represented by LBFs : $E_{\theta}(\theta)$.  
Projecting the first equation on the Fourier basis yields, under matrix form : 
\begin{equation}\label{eq12} 
\dfrac{k_i}{\varepsilon_i} J'_{i} c = \dfrac{k_o}{\varepsilon_o} \left( J'_{o} a +H^{+'} b \right)  
\end{equation}
Then projecting the second and third equations on the same basis gives :
\begin{equation}\label{eq13} 
J_i  c - J_o a - H^+ b   = \sigma G p  
\end{equation}
and 
\begin{equation}\label{eq14} 
\dfrac{-i Z_0}{n_i} J'_i c = [G , S] [p, q]^t
\end{equation}
Where matrices $G$ and $S$ are given by  $G_{nm} = \langle g_m(\theta), e^{in\theta} \rangle$ and  $S_{nm} = \langle s_m(\theta), e^{in\theta} \rangle$ , and more specifically: 
$$\begin{cases}
G_{mn} = \dfrac{-i{\theta_g}}{4\pi} e^{-i \frac{n \theta_g}{2}} \left\{ e^{i\frac{m \pi}{2}} \sinc(\alpha_{nm}^-)- e^{-i\frac{m \pi}{2}} \sinc(\alpha_{nm}^+) \right\} \\
S_{np}=\dfrac{1}{4} e^{-i n\theta_c}\left\{ e^{i\frac{p \pi}{2}} J_0(\beta_{np}^-) + e^{-i\frac{p \pi}{2}} J_0(\beta_{np}^+) \right\}  
\end{cases}$$
$[p,q]^t$ is the column vector gathering the coefficients $p_m$ and $q_m$, $\alpha_{nm}^\pm= (m\pi\pm n\theta_g)/2$ and $ \beta^\pm = (m\pi\pm n\bar{\theta}_g)/2$. Here we adopt the notation $\sinc(x)=\sin(x)/x$.  \\

Solving these algebraic systems leads to expressions of the coefficients $b$ and $c$ that have, once again, the same form as in equation \ref{eq5} where matrices $A$ and $B$ are given by:  
\begin{equation} \left\{
\begin{array}{ll}\label{eq15} 
  A = n_i J_{i}J'_{o} - n_oJ_oJ'_i +iZ_0\sigma J'_i W J'_o   \\    B =  n_oJ'_iH^+  - n_i J_iH^{+'} -i Z_0\sigma J'_i W H^{+'} 
  \end{array}\right.  \end{equation}
$W = [G, \mathbf{ 0}][G,S]^{-1}$  where $\mathbf{ 0}$ denotes the zero matrix having the size of $S$. Then coefficients $p$ and $q$ can be obtained simply from equation \ref{eq14}. \
\section{Results}\label{III}
\subsection{Convergence and stability}
Let us now compare the performances of the three approaches in terms of convergence and stability when the total number of Fourier harmonics retained in the numerical computations is increased. Such a number will be denoted by $N=2M+1$ where $M$ is, usually, called the truncation order. We consider a cylinder, with radius $R=\text{0.5}\mu\text{m}$, lying in vacuum and filled with a dielectric of relative permittivity $\varepsilon_i=\text{3.9}$, it is covered with a sheet of graphene whose parameters are $\theta_g=\pi$, $\mu=\text{0.5 eV}$, $\hbar \Gamma=\text{0.1 eV}$ taken at $T=300\text{K}$. The angle of incidence is first fixed to $\varphi=\pi/2$. \\
\begin{figure}[ht!]
{\includegraphics[width=8.cm]{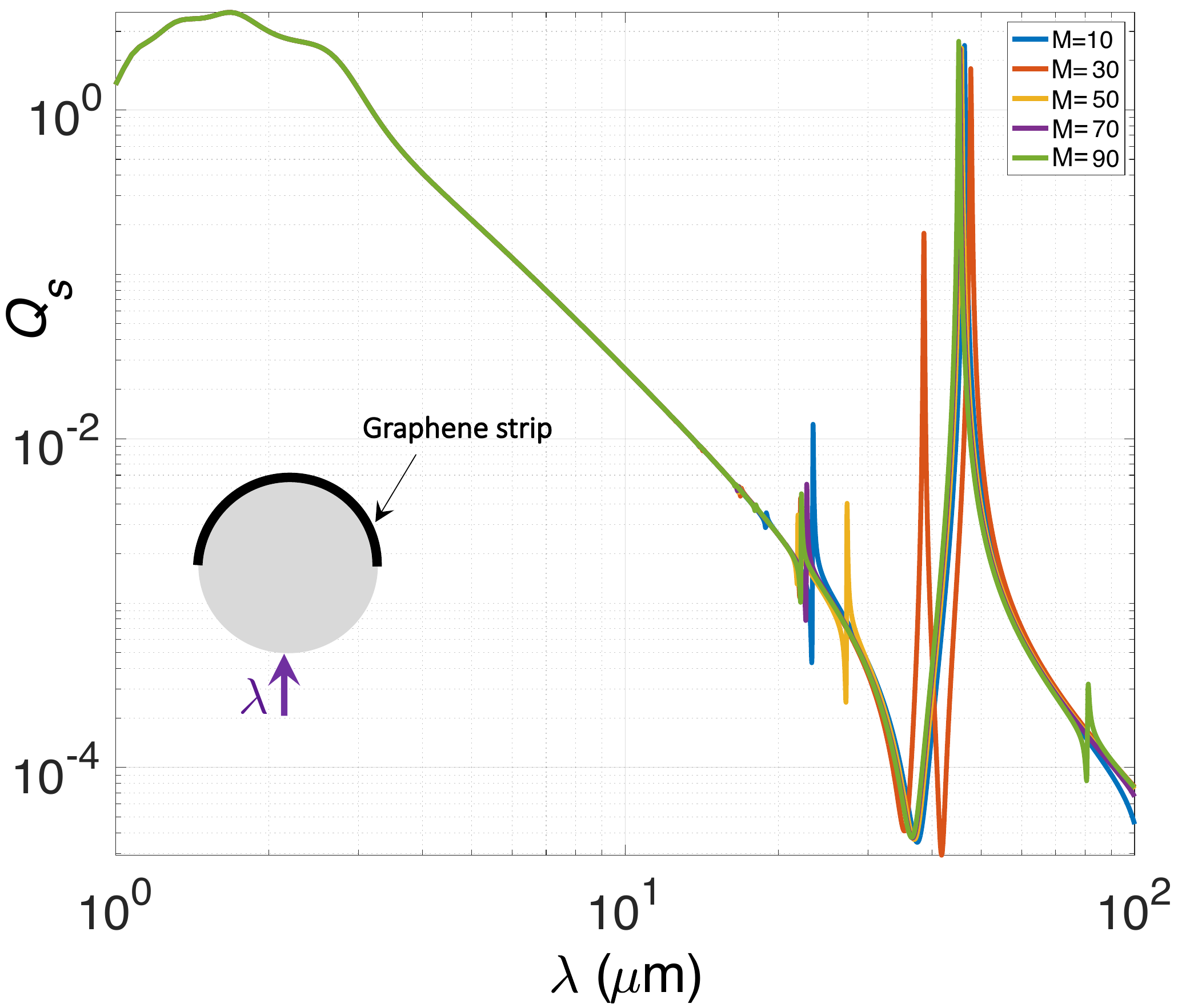}}
\caption{Scattering efficiency spectrum computed, with the classical FMM, for different values of the truncation order $M$.}
\label{Figure3}
\end{figure}
\begin{figure}
\centering
{\includegraphics[width=8.5cm]{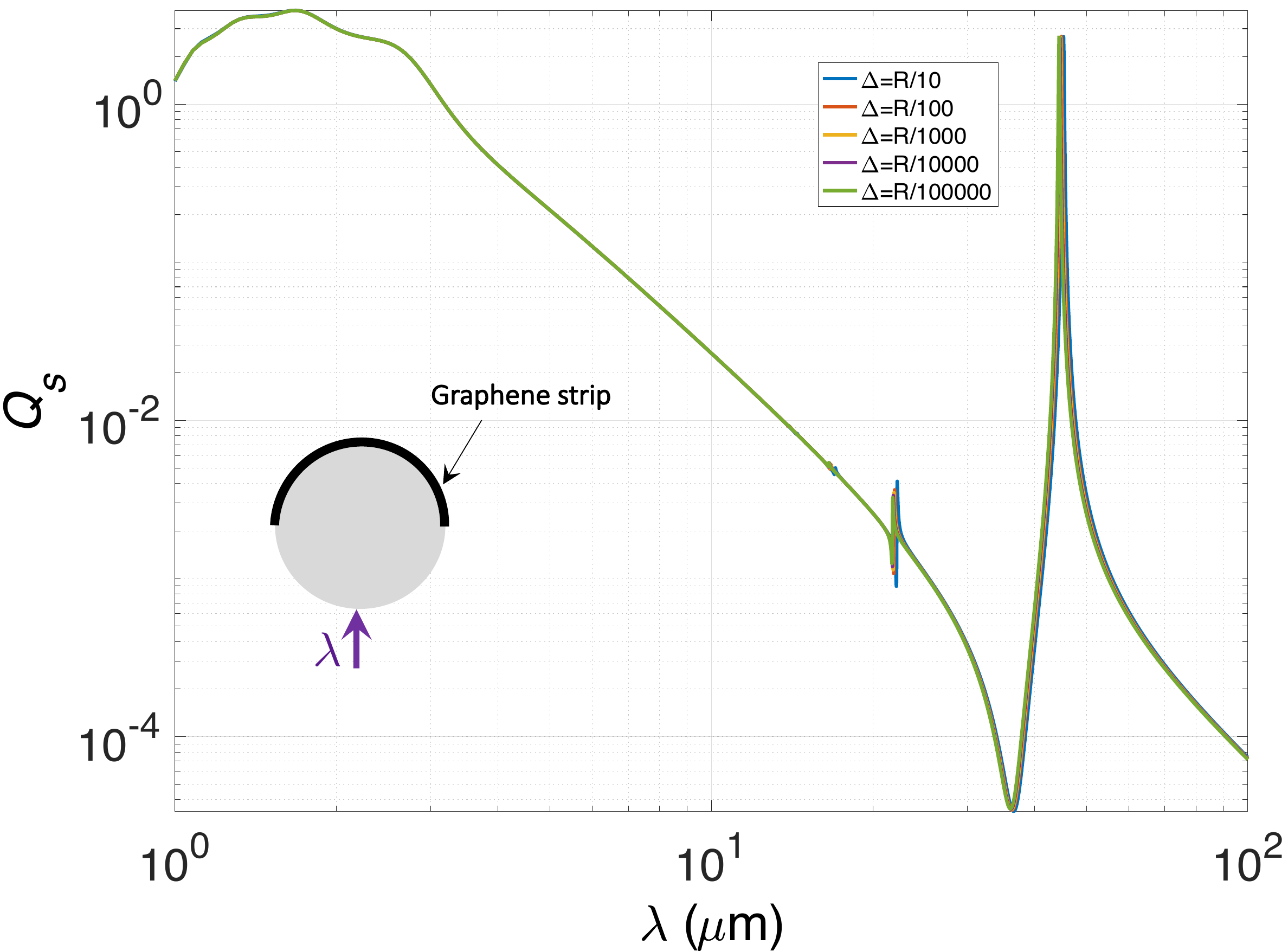}}
\caption{Scattering efficiency spectrum computed, with the FMM-ABC, for different values of $\Delta$ and for a truncation order : $M=90$.}
\label{Figure4}
\end{figure}
Figure \ref{Figure3} shows the spectrum of the scattering efficiency computed with the classical FMM approach for different truncation orders. We can clearly distinguish two subdomains in this spectrum : a low wavelengths subdomain where the method seems to behave properly and a high wavelengths subdomain where there are some resonances (due to surface plasmons polaritons on graphene \cite{SR_Cyl}) and where the method fails to converge. This manifests as many spurious and unstable peaks and dips appearing in this region of the spectrum. It is of fundamental importance to emphasize that increasing further the truncation order doesn't lead to any stabilization of these resonances. As stated in the theoretical section, this lack of convergence was expected for the classical FMM (it is due to the improper use of the correct Fourier factorization rules, cf. \citep{Li:FFF} and \cite{Khavasi}) and is clearly demonstrated in these calculations.

\begin{figure}
  \centering
  \begin{minipage}[b]{0.35\textwidth}
    \includegraphics[width=\textwidth]{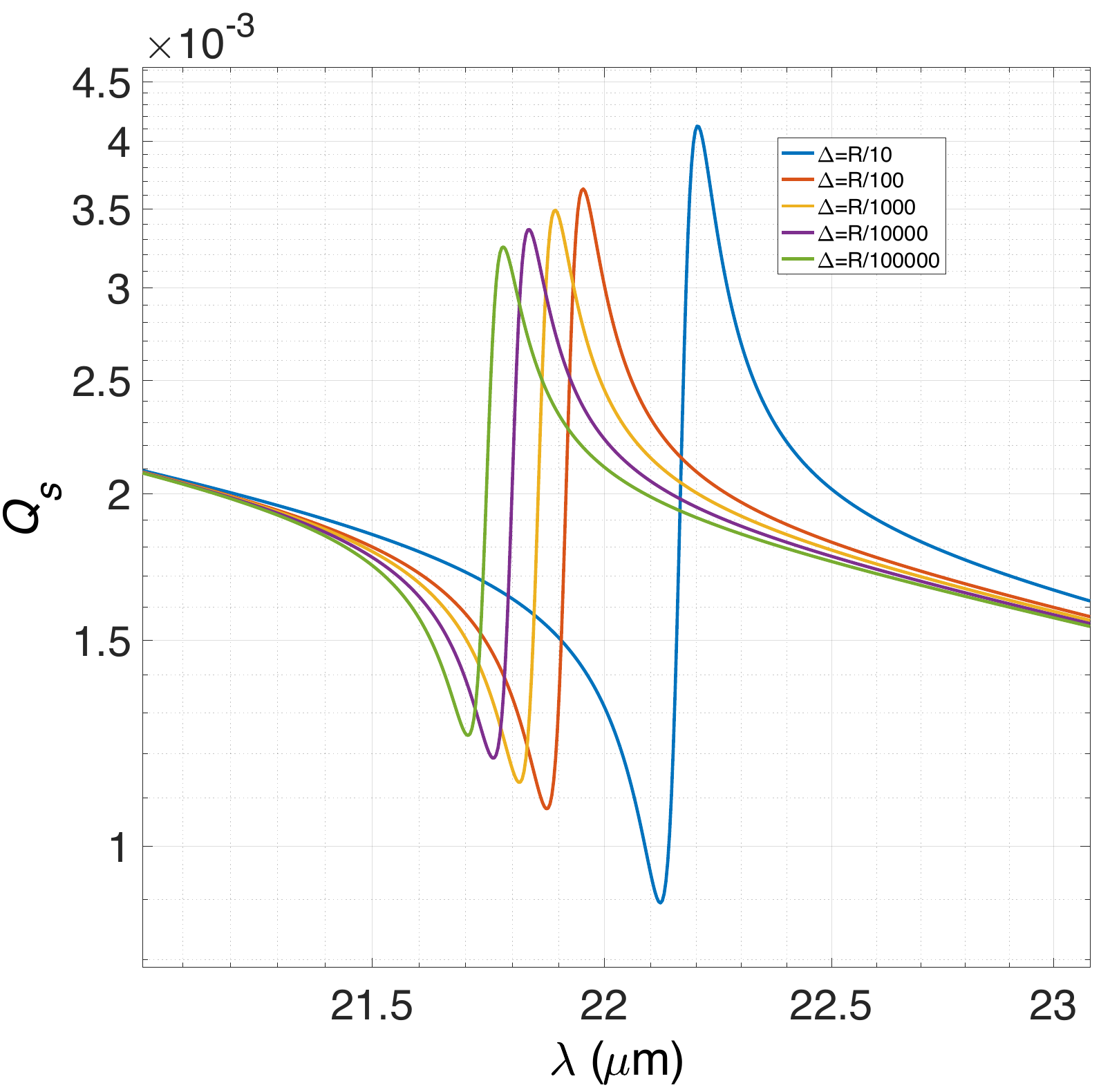}
  \end{minipage}
  \qquad
  \begin{minipage}[b]{0.35\textwidth}
    \includegraphics[width=\textwidth]{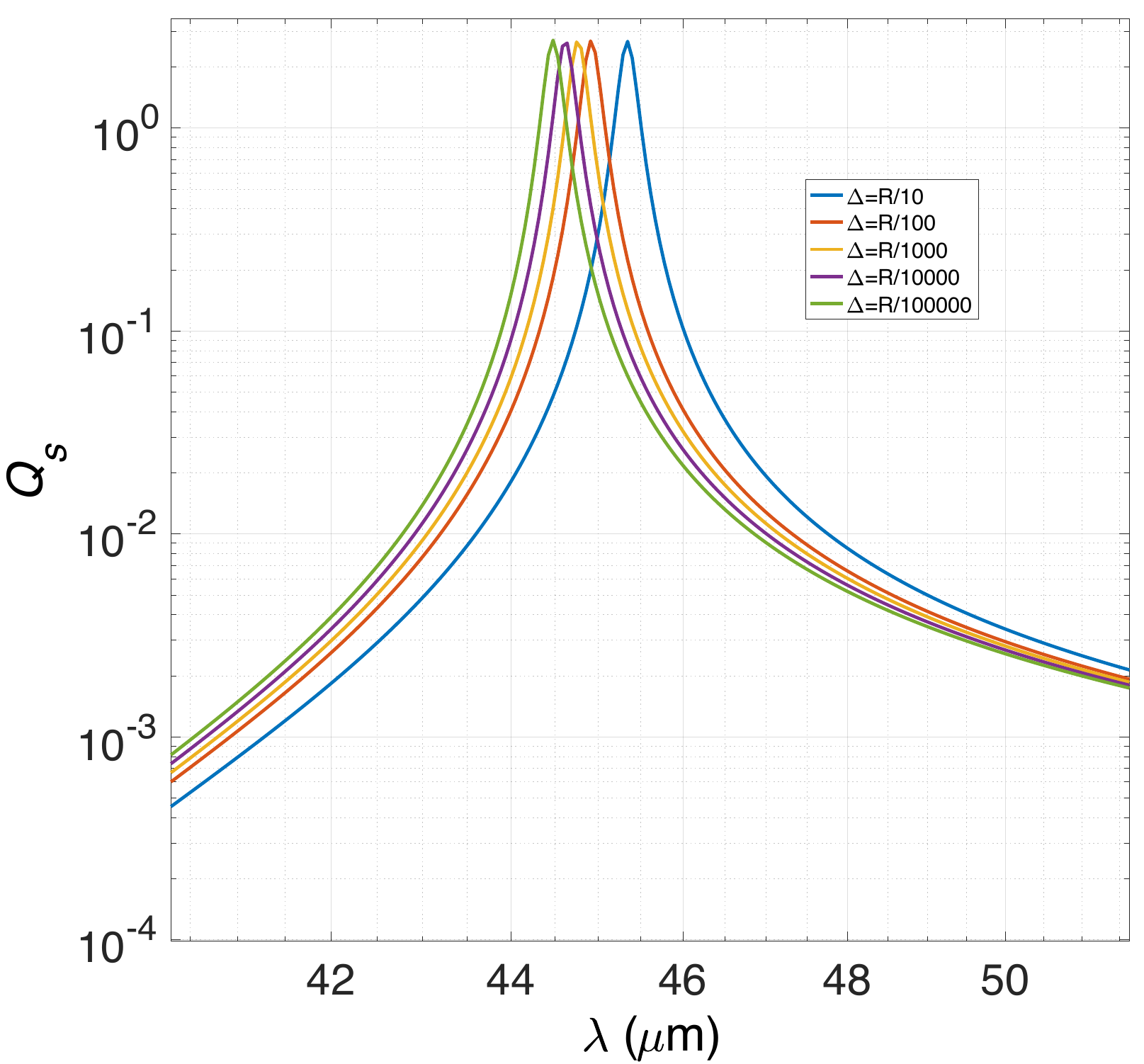}
  \end{minipage}
  \caption{Zooms over the two main resonances observed in figure \ref{Figure4}. Left around 22$\mu$m and right around 45$\mu$m.}
   \label{Figure5}
\end{figure}
\begin{figure}
\centering
{\includegraphics[width=8.5cm]{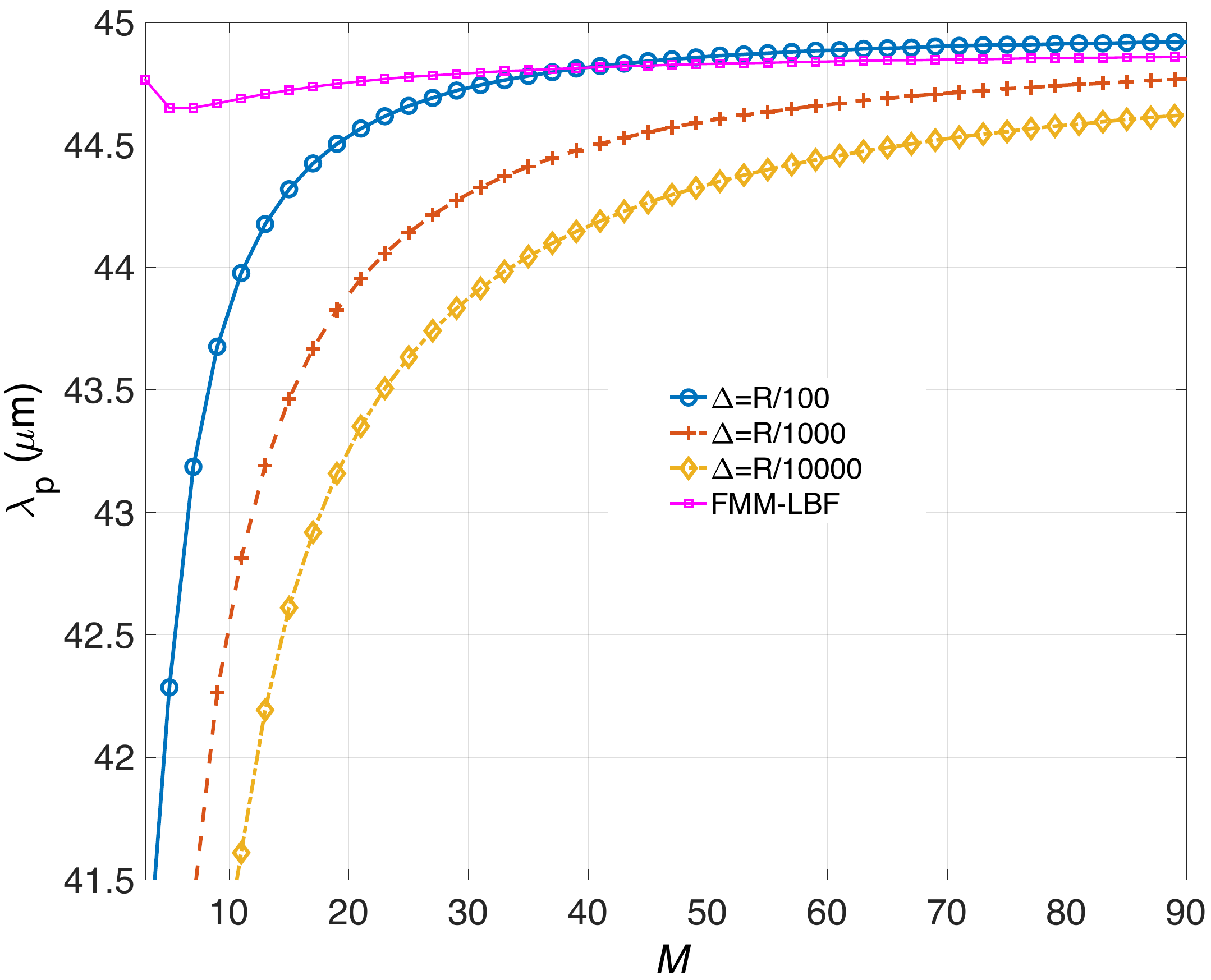}}
\caption{Evolution of the rightmost peak wavelength $\lambda_p$ versus $M$ for different values of $\Delta$. }
\label{Figure6}
\end{figure}
\begin{figure}
  \centering
  \begin{minipage}[b]{0.35\textwidth}
    \includegraphics[width=\textwidth]{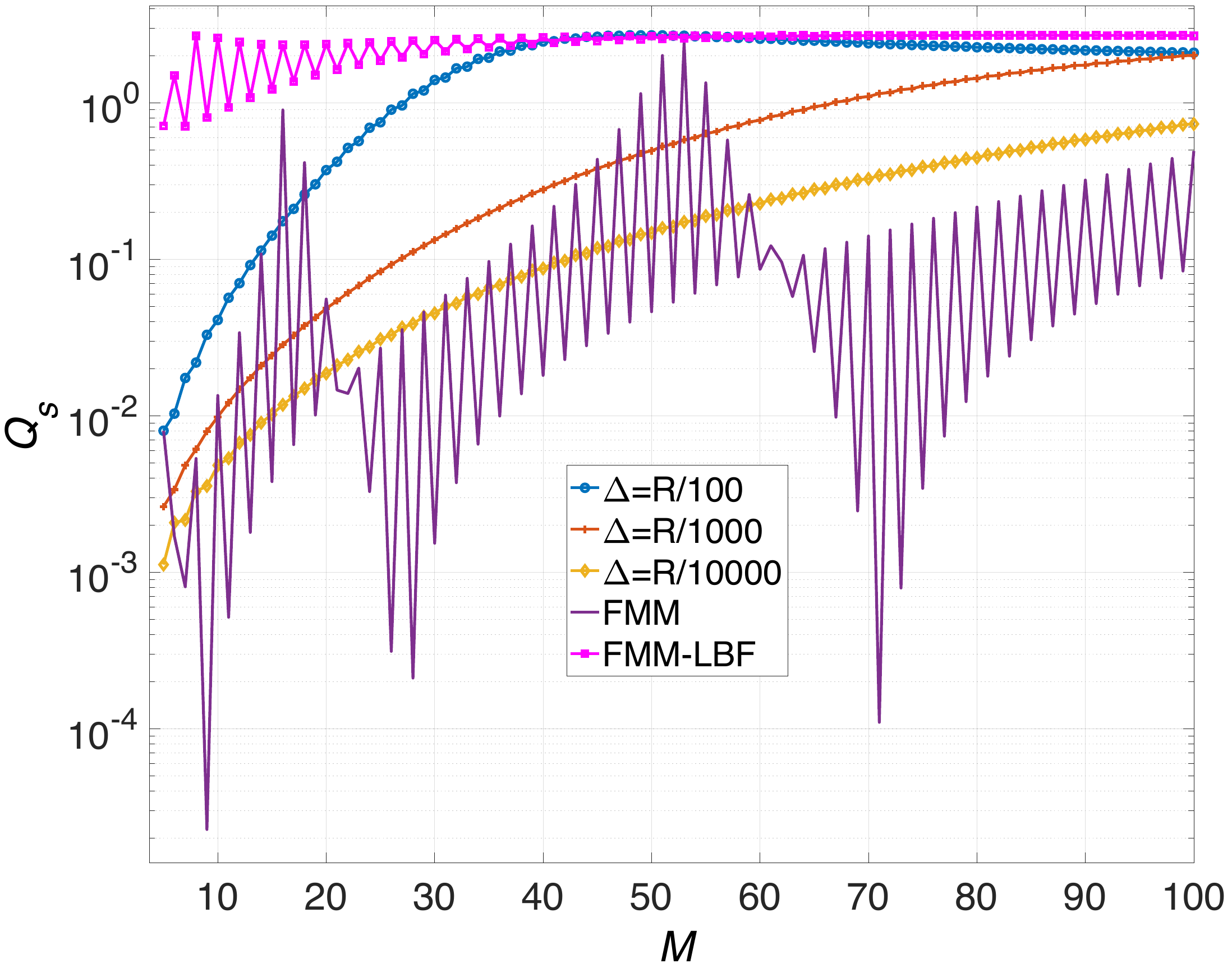}
  \end{minipage}
  \quad
  \begin{minipage}[b]{0.34\textwidth}
    \includegraphics[width=\textwidth]{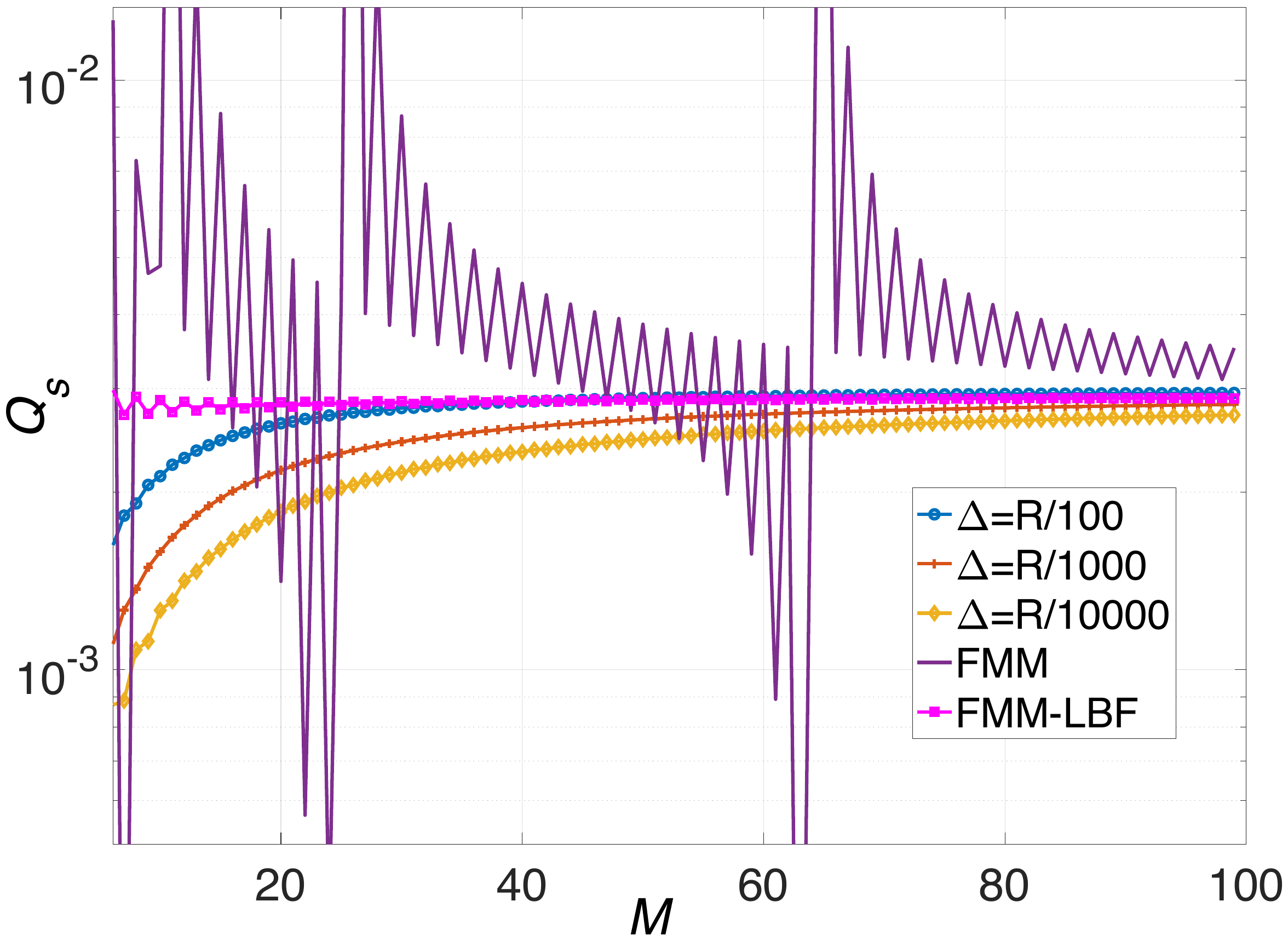}
  \end{minipage}
  \caption{$Q_s$ versus $M$ at resonance for $\lambda=44.86 \mu m$ (upper panel) and out of resonance for $\lambda=50 \mu m$ (lower panel). }
   \label{Figure7}
\end{figure}

Now, we are going to examine the behavior of the FMM with ABCs and explore the influence of the free parameter $\Delta$. Figure \ref{Figure4} shows the scattering efficiency spectrum computed for different values of $\Delta$ and for $M=90$, which is a rather high truncation order. First, we observe far better convergence and stability behaviours as compared to the classical FMM. Second, we clearly see the predominance of two main peaks keeping almost the same spectral positions, in contrast with the former results. A closer look at these peaks, reported in figure \ref{Figure5}, reveals that they shift towards low wavelengths, as the parameter $\Delta$ is decreased, and seem to converge to specific positions. To clarify this last point we focus on the rightmost peak and follow the evolution of its wavelength $\lambda_p$ as the parameters $M$ and $\Delta$ are varied. The results are shown in figure \ref{Figure6} where we plot $\lambda_p$ versus $M$ for three different values of $\Delta$. For each value of $\Delta$ we observe a convergence process leading to different values of $\lambda_p$. For comparison we added, in figure \ref{Figure6}, the results obtained by use of the FFM-LBF (magenta solid line with squares) which show a clear and fast convergence to $\lambda_p \simeq 44.86 \mu m$ which we will consider as the reference value \cite{Taiwan_LBF}. 
Thus we see that the FMM-ABC can predict quite correctly the shape and the number of resonances, but presets stability problems regarding the peaks associated to resonant phenomena in the structure. This is linked to the difficulty of choosing the right $\Delta$, all the more so that this latter may vary with the wavelength. 
For the FMM-LBF, on the contrary, there is no supplementary parameters involved and the method proves to be stable against the positions of sharp resonances. It is worth to emphasize that, from our numerical experiments, the same conclusion holds for the lower wavelength peak ($\lambda_p \simeq 21.93 \mu m$).  From the standpoint of convergence, the superiority of the FMM-LBF is evident as can be seen from figure \ref{Figure7} where we plot the scattering efficiency versus the truncation order $M$ for two wavelengths (at resonance and out of resonance) computed with the classical FMM, the FMM-ABC and the FMM-LBF. The FMM (solid purple line) is completely out of convergence for both cases while the FMM-ABC behaves slightly better but still presents the problem of choice of $\Delta$. The FMM-LBF shows clear convergence for both wavelengths though its is faster for the out of resonance case. Therefore, and because of its outstanding performances, the FMM-LBF will be our tool for investigating the physics behind the observed resonances in the considered structure.  \\
    
\subsection{Plasmonic modes over the graphene strip}
\begin{figure}[ht!]
\centering
{\includegraphics[width=8.5cm]{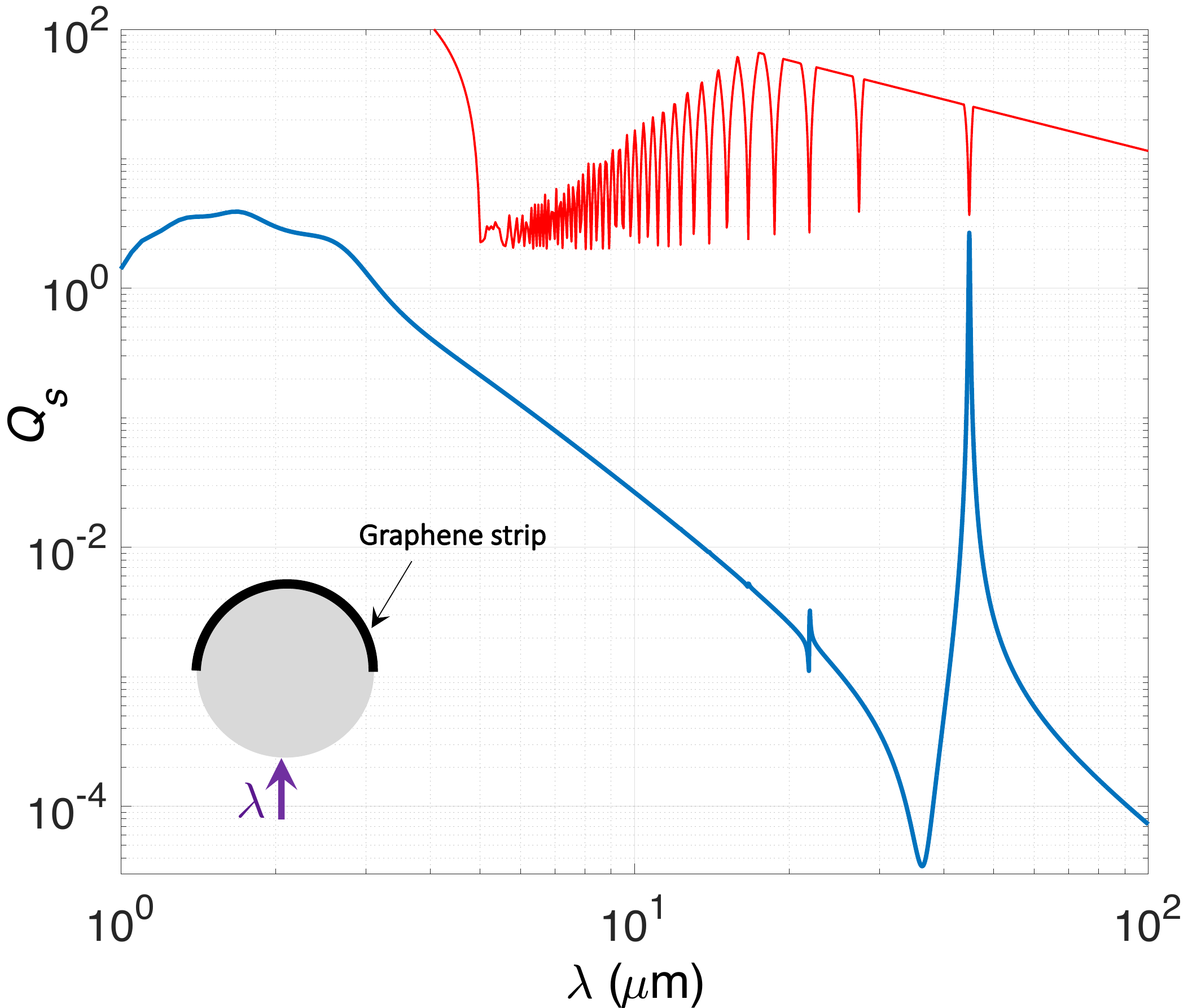}}
\caption{Scattering efficiency computed with the FMM-LBF (blue) for $M=90$. The red curve is represents the modulus of the smallest eigenvalue of matrix $B$ of equation \ref{eq15} (rescaled for the clarity of the figure).}
\label{Figure8}
\end{figure}

\begin{figure}[ht!]
\centering
{\includegraphics[width=8.5cm]{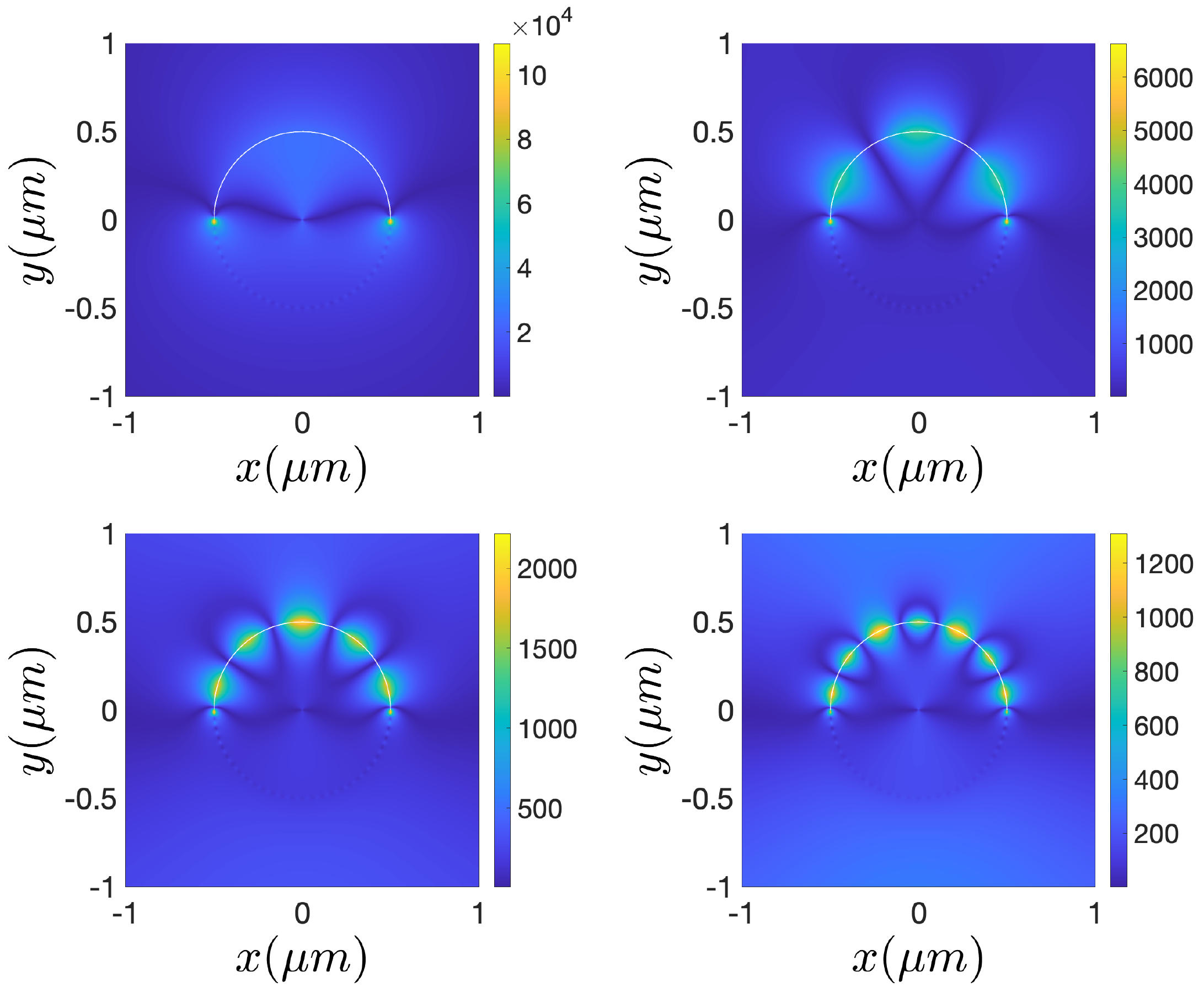}}
\caption{Map of the modulus of the electric component $E_{\theta}(x,y)$ computed for the four first resonant peaks observed in the spectrum of figure \ref{Figure8} : $ \lambda_1 = 44.86 \mu m$, $\lambda_3 =  21.93 \mu m$, $\lambda_5 = 16.66 \mu m$ and $ \lambda_7 = 13.98 \mu m$.}
\label{Figure9}
\end{figure}

\begin{figure}[ht!]
\centering
{\includegraphics[width=9cm]{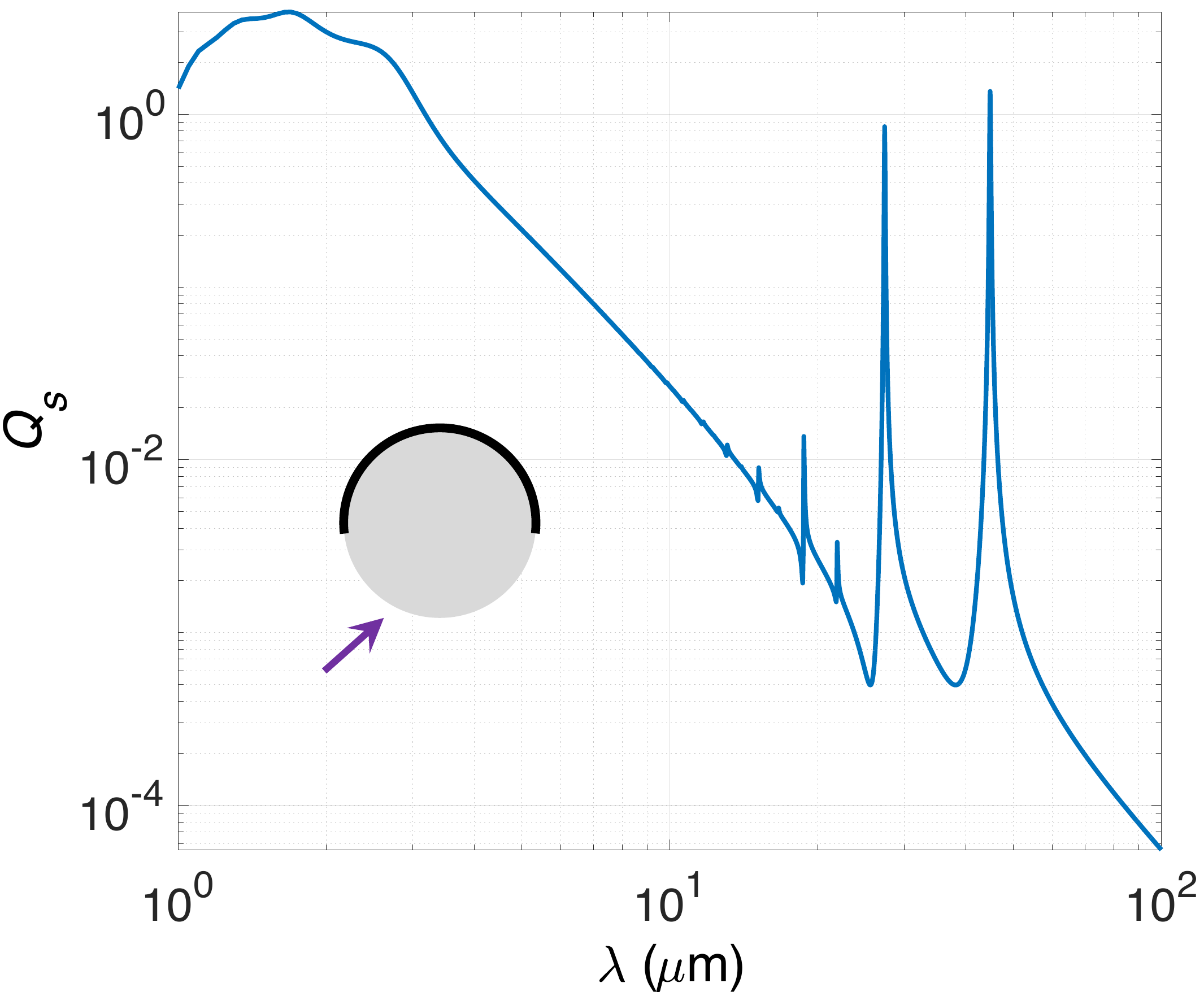}}
\caption{Scattering efficiency spectrum computed, with the FMM-LBF, at an incidence $\varphi = \pi/4$ with $M = 90$, the other parameters are unchanged.}
\label{Figure10}
\end{figure}

\begin{figure}[ht!]
\centering
{\includegraphics[width=9cm]{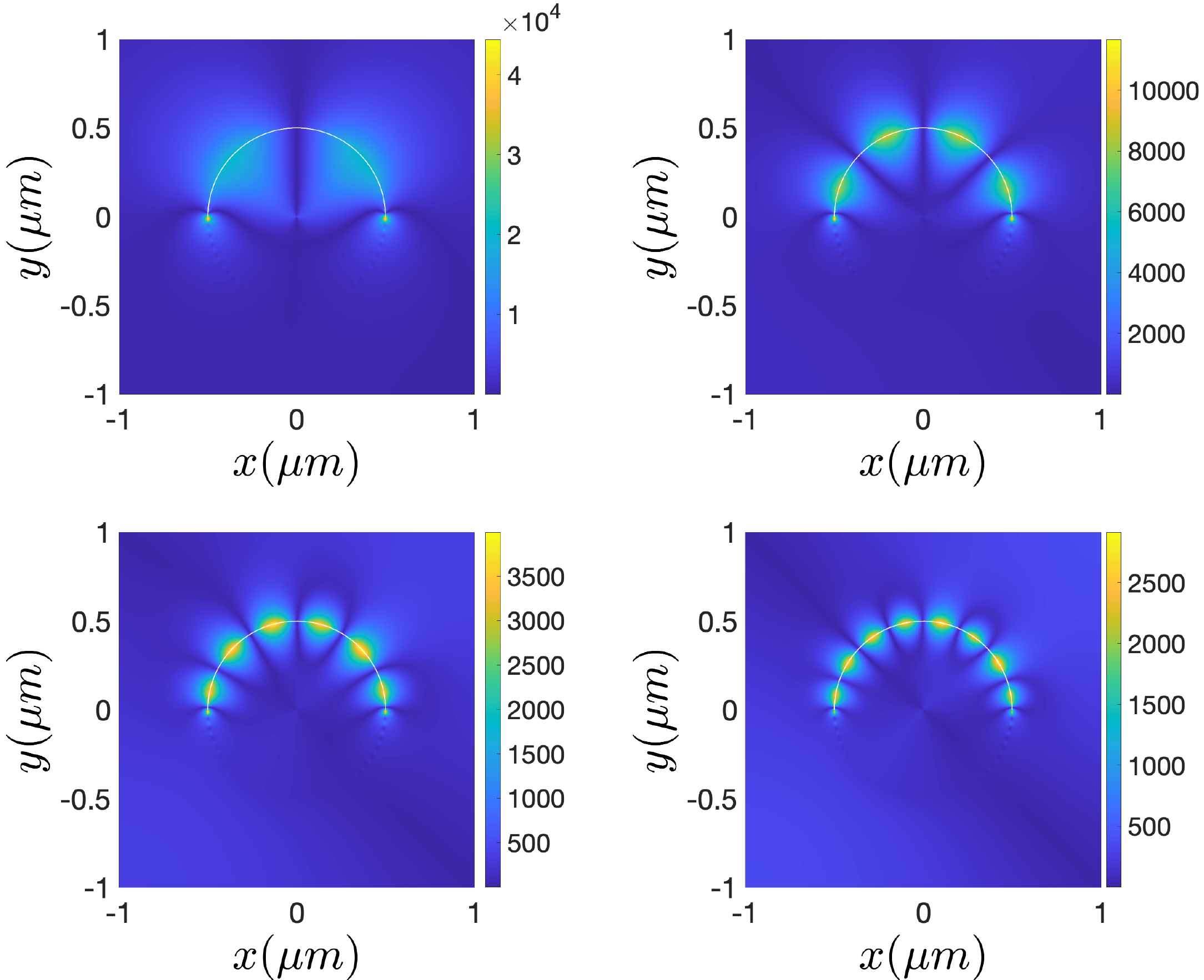}}
\caption{Map of the modulus of the electric component $E_{\theta}(x,y)$ computed for the new resonant peaks appearing in the spectrum of figure \ref{Figure10} : $ \lambda_2 = 27.34 \mu m$, $\lambda_4 = 18.72 \mu m$, $\lambda_6 = 15.16 \mu m$ and $ \lambda_8 = 13.07 \mu m$.}
\label{Figure11}
\end{figure}

We would like, now, to examine the physical origin of the resonances observed in the scattering efficiency spectrum. For that, we first compute (using the FMM-LBF) and report in figure \ref{Figure8}, the scattering efficiency spectrum together with the modulus of the smallest eigenvalue of matrix B of equation \ref{eq15} (rescaled for more clarity).
 The magnitude of this latter presents dips almost wherever the structure has a mode. We observe that there are more modes than what can be observed in the scattering efficiency spectrum. This is due to the lack of compatibility between the absent modes and the symmetry of the physical configuration. To gain more insight, we plot the map of the modulus of the electric component $E_{\theta}(x,y)$ around the coated cylinder for the four first resonant peaks observed in the spectrum : $ \lambda_1 = 44.86 \mu m$, $\lambda_3 =  21.93 \mu m$, $\lambda_5 = 16.66 \mu m$ and $ \lambda_7 = 13.98 \mu m$ (note that the last two peaks are tiny and hardly observable in figure \ref{Figure8}). The results are depicted in figure \ref{Figure9} where we first notice the expected singular behavior at the ends of the graphene sheet (namely at $\theta=0$ and $\theta=\pi$). The different peaks represent the plasmonic resonances \cite{SR_Cyl} supported by the graphene sheet at these wavelengths and it is of fundamental importance to remark that they, all, have an odd number of anti-nodes (1, 3, 5 and 7, hence their numbering) because of the symmetry of the physical configuration. The other plasmonic resonances (with an even number of anti-nodes) can be excited if one breaks this symmetry. An example of such a situation is given in figure \ref{Figure10} where we took an angle of incidence $\varphi=\pi/4$ and computed a new spectrum clearly revealing the missing plasmonic peaks. To complete the picture, we report in figure \ref{Figure11} the map of the modulus of the electric component $E_{\theta}(x,y)$ for the peaks $\lambda_2 = 27.34 \mu m$, $\lambda_4 = 18.72 \mu m$, $\lambda_6 = 15.16 \mu m$ and $ \lambda_8 = 13.07 \mu m$.


\section{Conclusion}
We presented an efficient and numerically very robust approach for the modelling of EM scattering from a cylinder partially covered with graphene. It is based on the classical Fourier-Bessel decomposition and classical boundary conditions to which we adjoint an {\it ad-hoc} expansion of the tangential electric field valid over the circumference. This latter is introduced to better take into account the singular nature of this field at the ends of the graphene sheet. Using this method we explored the scattering efficiency spectra of such structures and and showed that they present several peaks that we related to the existence of surface plasmons polaritons modes over graphene. The nature of each mode has been examined through the computation of the near field map of the electric field that revealed their standing wave nature. This method can be, safely, used in the study of the properties of this kind of setups exhibiting superscattering or invisibility for example. Finally it of interest to stress that it can be easily extended to multilayered cylinders with many strips at each interface which will allow a much more richer behavior.       
\vspace*{10cm}

 
\bibliographystyle{unsrt}


\begin{thebibliography}{10}

\bibitem{book1}
J.A. Kong.
\newblock {\em Electromagnetic Wave Theory}.
\newblock A Wiley-Interscience publication. Wiley, 1986.

\bibitem{Shanhui}
Zhichao Ruan and Shanhui Fan.
\newblock Superscattering of light from subwavelength nanostructures.
\newblock {\em Phys. Rev. Lett.}, 105:013901, Jun 2010.

\bibitem{CompactSS}
Rujiang Li, Bin Zheng, Xiao Lin, Ran Hao, Shisheng Lin, Wenyan Yin, Erping Li,
  and Hongsheng Chen.
\newblock Design of ultracompact graphene-based superscatterers.
\newblock {\em IEEE Journal of Selected Topics in Quantum Electronics},
  23(1):130--137, 2017.

\bibitem{GuizalPRE}
B.~Guizal and D.~Felbacq.
\newblock Numerical computation of the scattering matrix of an electromagnetic
  resonator.
\newblock {\em Phys. Rev. E}, 66:026602, Aug 2002.

\bibitem{Integral1}
Sergii~V. Dukhopelnykov, Ronan Sauleau, Maria Garcia-Vigueras, and Alexander~I.
  Nosich.
\newblock Combined plasmon-resonance and photonic-jet effect in the thz wave
  scattering by dielectric rod decorated with graphene strip.
\newblock {\em Journal of Applied Physics}, 126(023104):1--8, 2019.

\bibitem{Integral2}
Sergii~V. Dukhopelnykov, Ronan Sauleau, and Alexander~I. Nosich.
\newblock Integral equation analysis of terahertz backscattering from circular
  dielectric rod with partial graphene cover.
\newblock {\em IEEE Journal of Quantum Electronics}, 56(6):1--8, 2020.

\bibitem{GUIZAL1999}
B.~Guizal and D.~Felbacq.
\newblock Electromagnetic beam diffraction by a finite strip grating.
\newblock {\em Optics Communications}, 165(1):1--6, 1999.

\bibitem{Khavasi}
Amin Khavasi.
\newblock Fast convergent fourier modal method for the analysis of periodic
  arrays of graphene ribbons.
\newblock {\em Opt. Lett.}, 38(16):3009--3012, Aug 2013.

\bibitem{Li:FFF}
Lifeng Li.
\newblock Use of fourier series in the analysis of discontinuous periodic
  structures.
\newblock {\em J. Opt. Soc. Am. A}, 13(9):1870--1876, Sep 1996.

\bibitem{Taiwan_LBF}
Ruey-Bing Hwang.
\newblock Highly improved convergence approach incorporating edge conditions
  for scattering analysis of graphene gratings.
\newblock {\em Scientific Reports}, 10(1):12855, 2020.

\bibitem{Graphene1}
L~A Falkovsky.
\newblock {\em J. Phys.: Conf. Ser}, 129:012004, 2008.

\bibitem{GranetGuizal}
G.~Granet and B.~Guizal.
\newblock Efficient implementation of the coupled-wave method for metallic
  lamellar gratings in tm polarization.
\newblock {\em J. Opt. Soc. Am. A}, 13(5):1019--1023, May 1996.

\bibitem{SR_Cyl}
Mahin Naserpour, Carlos~J. Zapata-Rodr{\'\i}guez, Slobodan~M. Vukovi{\'c},
  Hamid Pashaeiadl, and Milivoj~R. Beli{\'c}.
\newblock Tunable invisibility cloaking by using isolated graphene-coated
  nanowires and dimers.
\newblock {\em Scientific Reports}, 7(1):12186, 2017.

\end{thebibliography}
\end{document}